\pdfoutput=1
\documentclass[a4paper]{article}
\usepackage{amsmath,graphicx,amssymb,xcolor,hyperref}
\usepackage[flushleft]{threeparttable}
\usepackage{INTERSPEECH2021}
\usepackage{array,mathtools}
\newcolumntype{P}[1]{>{\centering\arraybackslash}p{#1}}

\title{Improving Perceptual Quality by Phone-Fortified Perceptual Loss using Wasserstein Distance for Speech Enhancement}
\name{Tsun-An Hsieh$^{1}$ , Cheng Yu$^{1}$, Szu-Wei Fu$^{1}$ , Xugang Lu$^{2}$, and Yu Tsao$^{1}$}
\address{
  $^{1}$Research Center for Information Technology Innovation, Academia Sinica, Taiwan \\
$^{2}$National Institute of Information and Communications Technology, Japan}
\email{\{tahsieh, chengyu\_citi, jasonfu, yu.tsao\}@citi.sinica.edu.tw, xugang.lu@nict.go.jp}

\begin{document}

\maketitle
\begin{abstract}
Speech enhancement (SE) aims to improve speech quality and intelligibility, which are both related to a smooth transition in speech segments that may carry linguistic information, e.g. phones and syllables. In this study, we propose a novel phone-fortified perceptual loss (PFPL) that takes phonetic information into account for training SE models. To effectively incorporate the phonetic information, the PFPL is computed based on latent representations of the \textit{wav2vec} model, a powerful self-supervised encoder that renders rich
phonetic information. To more accurately measure the distribution distances of the latent representations, the PFPL adopts the Wasserstein distance as the distance measure. Our experimental results first reveal that the PFPL is more correlated with the perceptual evaluation metrics, as compared to signal-level losses. Moreover, the results showed that the PFPL can enable a deep complex U-Net SE model to achieve highly competitive performance in terms of standardized quality and intelligibility evaluations on the Voice Bank--DEMAND dataset.
\end{abstract}

\section{Introduction}

In real-world speech-related applications, speech signals may be contaminated by environmental noise, and thus constrain the achievable performance on target tasks. To address this issue, speech enhancement (SE) has been studied for decades. Numerous signal processing-based methods \cite{boll1979suppression, lim1979enhancement, paliwal1987speech, loizou2013speech} have been proposed. 
% These methods are based on the assumed statistical properties of speech and noise signals. Unfortunately, when these assumed properties are unfulfilled, SE performance may drop drastically.
These methods are based on the assumed statistical properties of speech and noise signals. Unfortunately, SE performance may drop drastically when these assumptions are unfulfilled.
With recent advances in neural network (NN) models, SE performance has increased notably. Well-known NN models, such as deep denoising autoencoder (DDAE) \cite{lu2013speech}, deep neural networks (DNNs) \cite{xu2014regression}, recurrent neural networks (RNNs) \cite{weninger2014single}, long short-term memory (LSTM) \cite{weninger2015speech}, convolutional neural networks (CNNs) \cite{zhao2018convolutional}, fully convolutional networks (FCNs) \cite{fu2018end, pandey2019tcnn}, convolutional recurrent neural networks (CRNNs) \cite{tan2018convolutional}, and generative adversarial networks (GANs) \cite{pascual2017segan, soni2018time, pandey2018adversarial, baby2019sergan, qin2018improved, su2020hifi, fu2019metricgan} have made notable improvements over traditional signal processing-based SE methods. \par

For these NN-based SE approaches, designing a suitable objective function is a crucial factor. Traditionally, point-wise distances are often used to form the objective functions. Point-wise distances, such as $L^1$ and/or $L^2$ norms between paired noisy-clean speech signals, attempt to recover information on a signal level. Recent studies have revealed that objective functions based on point-wise distances may not fully reflect the perceptual difference between noisy and clean speech signals. As the purpose of SE is to recover speech quality and intelligibility, objective functions that consider perceptual metrics have been investigated for NN-based SE. In some studies, perceptual metrics were modified to their differentiable alternatives for convenient gradient calculations to optimize the NN parameters. Some notable works are the perceptual evaluation-based loss function \cite{martin2018deep}, joint source-to-distortion ratio (SDR) perceptual evaluation for speech quality optimization \cite{kim2019end}, and modified short-time objective intelligibility (STOI) loss functions for network optimization \cite{fu2018end, kolbaek2018monaural, zhao2018perceptually}. Along this line, several studies focus on training NN models with target metrics for SE tasks \cite{fu2019learning}, as well as with GAN approaches like HiFi-GAN \cite{su2020hifi} and MetricGAN \cite{fu2019metricgan}.
%Other than direct optimizations for evaluation metrics, objective functions can be designed to minimize loss based on representations in latent spaces, where the latent spaces are from a pre-trained model that is given paired noisy-clean speech signals. 
% \textcolor{red}{Other than direct optimizations toward evaluation metrics, objective functions can be designed to minimize loss based on representations in latent spaces.} 
Another class of approaches focused on building loss functions in the spaces mapped by certain pre-trained classifiers. 
For example, in style transfer studies of computer vision, \cite{johnson2016perceptual} proposed training feed-forward networks based on perceptual loss. In \cite{germain2019speech}, the authors proposed utilizing an acoustic scene (AS) recognition network's latent spaces for the loss function, termed deep feature loss (DFL). For further improvement over DFL, \cite{kataria2020perceptual} proposed the perceptual ensemble regularization loss (PERL) as a variant of DFL that gathers several pre-trained models related to speech or acoustic tasks, achieving state-of-the-art performance in terms of quality. Despite the success, it remains unclear how acoustic event (AE) or AS classifiers benefit SE. \par

%In this paper, apart from using AE/AS classifiers, we show that phonetic characteristics are essential for SE. Therefore, we propose a specialized phone-fortified perceptual loss (PFPL), and elucidate how the it provides gradients that help SE training.
% \textcolor{red}{In this paper, we show that phonetic information is essential for SE. Therefore, apart from using AE/AS classifiers, we propose a specialized phone-fortified perceptual loss (PFPL), and elucidate how SE training can benefit from the discriminative capability of PFPL between the encoded noisy and clean features' distributions.}

In this paper, we propose a novel phone-fortified perceptual loss (PFPL) for training SE models. The PFPL modifies the original DFL in two aspects. First, the PFPL intends to consider the phonetic information embedded in the speech signals. Therefore, rather than using the AS recognition models, the PFPL is computed based on the latent representations of the \textit{wav2vec}
 model \cite{schneider2019wav2vec}, a powerful self-supervised encoder that renders rich phonetic information.  Second, as the distance used in the original DFL ignores the geometry of the distributions of the latent representations, PFPL adopts the Wasserstein distance \cite{OLKIN1982257}
as the distance measure. In this way, the SE training can be seen as an optimal transport problem that transforms the distributions of noisy speech to that of clean speech.
%Experimental reencodesults have shown that phonetic characteristics do benefit SE, and the proposed extension using Wasserstein distance yields better performance in quality and intelligibility.
Experimental results first confirmed that the PFPL can enable a deep complex U-Net SE model to achieve highly competitive performance on the Voice Bank--DEMAND dataset. A series of ablation studies investigated the effectiveness of individual parts in the PFPL.
% [a little bit of PFP explain here should be enough.]
% The purpose for SE is to improve speech quality and intelligibility. As many perceptual experiments show \cite{}, quality and intelligibility are correlated to temporal context segments of speech which are tightly connected to linguistic level of speech signal, e.g. phones, syllables. However, in most SE algorithms, there is no explicit integration of the temporal context dependent information into consideration (some may implicitly consider the information). In this paper, we explicitly take SC (SC) into consideration for SE. Moreover, based on the SC, a phone fortified perceptual (PFP) loss is designed for SE model optimization based on noisy-clean speech pairs. 
% [a little bit of PFP explain here should be enough.]
\begin{figure*}[t]
    \centering
    \includegraphics[width=0.95\textwidth]{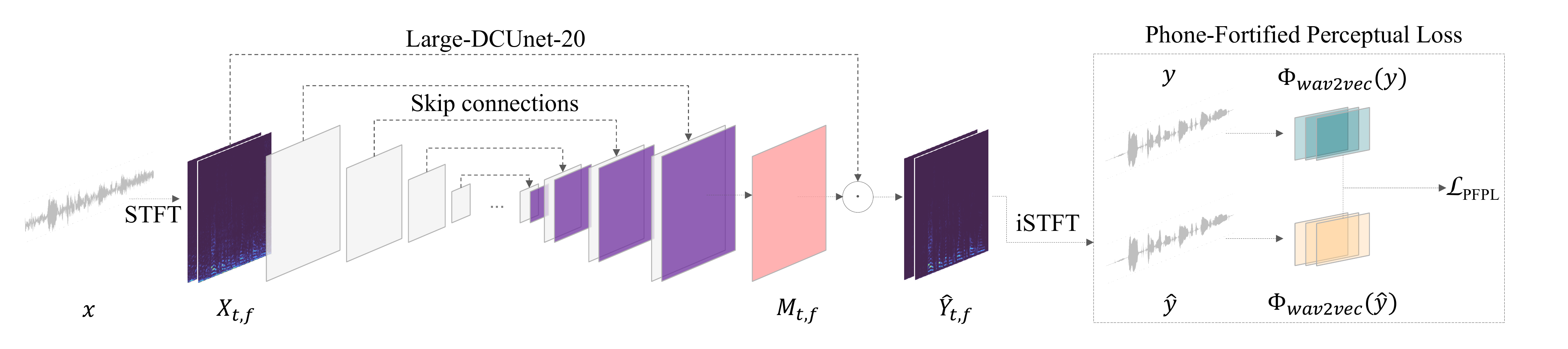}
    % \vspace{-0.3cm}
    \caption{A demonstration of the proposed network. The enhancement model estimates a cRM by the noisy spectrum, and consequently produces an enhanced spectrum. The PFPL then compares the semantic difference of clean speech and the enhanced one.}
    \label{fig:model}
    % \vspace{-0.4cm}
\end{figure*}

\section{Related Works}

\label{sec:relatedworks}
In this section, we first present DFL and PERL, which was mentioned in the previous section, with a more detailed discussion in Section \ref{ssec:dfl}. We then review the perceptual metrics approximated with trained networks. Such a network can work as a discriminator in a GAN or a stand-alone metric. Last but not least, we review methods that maximize the mutual information between contexts in Section \ref{ssec:cpc_w2v}. 

\subsection{DFL and PERL}
\label{ssec:dfl}
The idea to incorporate AS recognition in SE was first proposed in DFL \cite{germain2019speech}. According to \cite{su2020hifi}, the latent features from a pre-trained recognition network (used for machine perception) are used to approximate human perception (SE, in this case). Analogous to DFL, \cite{kataria2020perceptual} extend the idea to ensemble of six types of pre-trained networks, including AE classifiers and speech encoders. In spite of that PERL-AE uses AE classifier alone and yields the best result, it remains unexplainable due to the complication of characteristics in AEs, which are too difficult to be analyzed.
% However, acoustic scene recognition seems to be lacking in phonetic characteristic information, which we believe is the key to optimizing SE with respect to human perception. 

\subsection{MetricGAN and HiFi-GAN}
\label{ssec:gan}
 MetricGAN \cite{fu2019metricgan} applies a discriminator (also called Quality-Net \cite{fu2018quality}) to approximate the behavior of the evaluation functions of interest. The predicted score can also be treated as a special case of perceptual loss, with an embedding dimension equal to 1. Due to the limited dimensions, Quality-Net is easily fooled by the speech generated by the updated generator. Therefore, MetricGAN needs to alternatively train between the generator and the discriminator which slows down its efficiency. HiFi-GAN \cite{su2020hifi} incorporates the idea of GAN and deep feature loss. However, its deep feature loss is based on the discriminator, which may not be highly related to human perception.

\subsection{Contrastive Predictive Coding (CPC) and \textit{wav2vec}}
\label{ssec:cpc_w2v}
% 
%To make the best use of phonetic characteristics, we surveyed representation learning methods that could automatically discover representative features directly from raw data.
% \textcolor{red}{Recent studies of representation learning have shown capabilities in extracting representative features directly from raw data.}
%Some remarkable approaches have been proposed recently in this research field.
% \textcolor{red}{Some remarkable approaches are recently proposed in this research field.}
Recent studies of representation learning have shown capabilities in extracting representative features without supervision.
For instance, CPC \cite{oord2018representation} is an self-supervised method that proposes to extract task-agnostic features from high-dimensional data. It was the contrastive loss that helps capture features which maximize the amount of underlying shared information of the observation and its latent representation.
As a result, self-supervised methods that can extract features with phonetic information from speech signals drew our attention.
%Then, we focused on speech-related applications utilizing representation learning approaches.
We then focus on speech-related applications that utilizes representation learning approaches.
\par
The self-supervised automatic speech recognition (ASR) \textit{wav2vec} \cite{schneider2019wav2vec}, utilizing the CPC technique, has shown great performance in recognition accuracy and thus fits our interests. In practice, speech signals are first encoded with an encoder network that extracts features rich in phonetic information.
An ASR decoder is then trained based on these features as inputs.
%Based on our understanding in speech acoustics, phonetic characteristic can be greatly distorted due to noise contamination. \par
% 
\subsection{Wasserstein Distance}
\label{ssec:wasserstein}
The Wasserstein distance \cite{OLKIN1982257} is a measurement of two probability distributions on a metric space $(\mathcal{M}, d)$ with $d:\mathcal{M}\times\mathcal{M}\rightarrow\mathbb{R}$ a metric on $\mathcal{M}$. The Wasserstein distance of two densities $\mu$ and $\nu$ is defined as:
\begin{align}
\label{eq:wass}
    \mathcal{W}_p(\mu,\nu)\coloneqq
    \left(
    \inf_{\gamma\in\Gamma(\mu,\nu)}\int_{\mathcal{M}\times\mathcal{M}}d(x,y)^p{\rm d}\gamma(x,y)
    \right)^\frac{1}{p}
\end{align}
where $\Gamma(\mu, \nu)$ denotes a set of all possible measures (or {\it couplings}) on $\mathcal{M}\times\mathcal{M}$ with marginals $\mu$ and $\nu$. Here, $\gamma(x,y)$ is the coupling, that is, a joint distribution of marginals $\mu$ and $\nu$, representing any possible transport plan from $\mu$ to $\nu$. Comparing to $L^p$ distances that only regard the amount of mass transported, Eq. \eqref{eq:wass} shows that Wasserstein distance additionally takes the transport method into account. 
\section{Proposed framework}
\label{sec:proposedframework}
In this section, we start with introducing a complex U-Net adopted, which was widely utilized in several studies \cite{YaoA19coarse, hu2020dccrn, choi2018phase} that has been confirmed to achieve promising results. In the followings, we will describe the PFPL, which is a perceptual loss incorporated with Wasserstein distance in detail.

% Two aspects are important in designing an effective NN-based SE system. One is the model architecture, by which complex mapping functions between noisy and clean speech can be efficiently approximated. The other is the objective function, by which necessary speech information should be retained for SE during model optimization.
\subsection{Model Architecture}
\label{ssec:enhancement_model}

Inspired by the deep complex U-Net (DCUnet) in \cite{choi2018phase}, we designed a modified framework that estimates complex ratio masks (cRM) for a noisy complex spectrum with a different normalization mechanism. More specifically, as shown in Fig. \ref{fig:model}, a noisy speech signal is first converted to a complex spectrum through short-time Fourier transform (STFT), and the enhancement model generates a cRM. Subsequently, the noisy spectrum is multiplied by the cRM in a point-wise manner to derive the final enhanced spectrum, which is transformed to a waveform by inverse STFT (iSTFT). Here, according to \cite{choi2018phase}, a scheme that produces cRM with a complex neural network (cRM$\mathbb{C}$n) is used in this work, and we take the Large-DCUnet-20 as a reference architecture for our enhancement model. As a number of previous works \cite{ulyanov2016instance, conf/iccv/HuangB17} have indicated that instance normalization outperforms batch normalization on generation tasks by preserving the independence of samples in a mini-batch, we substitute the batch normalization layers with instance normalization layers. To describe the enhancement process precisely, given the noisy input speech signal $x$, the noisy spectrum $X_{t,f}$ is produced by STFT, such that $X_{t,f} = \mathrm{STFT}(x)$. Then, the enhancement model generates a cRM $M_{t,f}$ to produce the enhanced spectrum that $\hat{Y}_{t,f} = M_{t,f}\cdot X_{t,f}$, and transforms it to the enhanced waveform $\hat{y}$ by iSTFT.
\subsection{Phone-Fortified Perceptual Loss}
\label{ssec:problem_definition}
% \textcolor{purple}{To our instinct, it is essential that the objective functions for SE tasks are more focused on the integrity of speech enhanced with respect to the feature domain of phonetic characteristics (or phonetic characteristic level) rather than simply considering those at the signal level. That is, SE that aims to minimize signal-level losses does not guarantee integrity at the phonetic characteristic level, and thus suffers from observable mismatches with respect to perceptual metrics, despite satisfactory signal-level losses. As a result, our proposed PFPL adopts the latent space of the \textit{wav2vec} encoder network, which extracts phonetic characteristic-rich features. We then design a PFPL that performs distance calculation at phonetic characteristic level. We also incorporate mean absolute error (MAE) with PFPL to ensure optimization at both the phonetic characteristic level and at the signal level for SE.}\par

%\textcolor{blue}{In SE training, it is common to train an enhancement network using point-wise loss functions in either time domain or time-frequency domain. Despite that these approaches have achieved promising results, point-wise losses remain inconsistent with perceptual evaluations such as PESQ or STOI. Comparing to point-wise losses that measure distances in the signal level, the perceptual loss is devised to measure the distance in the latent space. In other words, the perceptual loss tends to encourage the estimates to be semantically similar to the targets \cite{johnson2016perceptual} rather than in signal level.
For training SE models, point-wise loss functions are commonly used in either time domain or time-frequency domain. Despite that these approaches have achieved promising results, point-wise losses remain numerically inconsistent with perceptual evaluations such as PESQ or STOI. Unlike point-wise losses that measure distances in the signal level, the perceptual loss is devised to measure the distance in the latent space \cite{johnson2016perceptual}. \par
% The perceptual loss has been proven to encourage the estimates to be semantically similar to its targets \cite{johnson2016perceptual}. \par
The design of perceptual losses requires an appropriate feature extractor. In the SE scenario, the estimates of a system are speech signals. It is thus our desire to carry out loss computation that is able to preserve attributes in speech signals (i.e., phones, speaker characteristics, etc.) in the training stage. Some proposed a supervised pre-trained encoder for loss computation, like \cite{germain2019speech}. However, these methods can suffer from the drawback of one-hot encoding in which the correlations between categories were ignored since the labels are in an orthogonal (high-dimensional) space \cite{rodriguez2018beyond}. As a consequence, the correlations between phones could be underestimated and thus restrict the capability of preserving attributes. Hence, we employ {\it wav2vec}, a self-supervised encoder, to compute the PFPL in a non-orthogonal (low-dimensional) space that is more capable of preserving attributes. Owing to the fact that speech signals carry linguistic information (e.g., phones, syllables, etc.) more often than noises do, we prefer to use models that generate features which are representative for phonetic information. Meanwhile, note that CNNs are known for the shift-invariance, which is analogous to the perceptual evaluation of speech quality (PESQ) \cite{PESQ} which is insensitive to shifts in a short-time. As stated above, we believe the CNN-based \textit{wav2vec} encoder (denoted as $\Phi_{\textit{wav2vec}}$) is suitable for the design of our loss function. \par
% For PFPL, the encoder $\Phi_{\textit{wav2vec}}$ of \textit{wav2vec large} transforms raw waveform inputs into a batch of sequence of 512 dimensional vectors. 
In contrast to previous works on perceptual loss that utilize activations in multiple layers, we merely extract the final outputs for efficiency. Formally, as the densities remain unknown, we define the PFPL by the Kantorovich--Rubinstein \cite{optimaltransport} dual form of Wasserstein distance:
\begin{align}
    \label{eq:loss}
    \mathcal{L}_{\mathrm{PFPL}}(y, \hat{y}) \coloneqq 
    \left\Vert y - \hat{y} \right\Vert_1+\sup_{f\in\mathcal{F}} 
    \mathbb{E}_{\mu}\left[f(c)\right] -
    \mathbb{E}_{\nu}\left[f(\hat{c})\right]
\end{align}
where $c=\Phi_{\textit{wav2vec}}(y)$ and $\hat{c}=\Phi_{\textit{wav2vec}}(\hat{y})$ are the features of the clean speech $y$ and the enhanced speech $\hat{y}$, respectively. Here, $\mu$ and $\nu$ are the densities of $c$ and $\hat{c}$ in the latent space, and $f$ is a function belonging to a set $\mathcal{F}: \{ f: \mathbb{R}^n \to \mathbb{R}^n | \|f(x_1) - f(x_2) \| \leq 1\|x_1 - x_2\|, \forall x_1 , x_2 \in \mathbb{R}^n \}$ of all 1-Lipschitz functions. For the given paired enhanced speech and clean speech, the PFPL minimizes the distance between the distributions of the estimates and the corresponding targets in a space of phonetic representations. Please note Eq.~(\ref{eq:loss}) that the PFPL includes an mean absolute error (MAE) loss to measure the signal-level difference. The effect of the MAE loss will be discussed in Section \ref{ssec:qualitative_analysis}. 
% 
%\section{Methodology}
%\label{sec:methodology}
% 
\begin{figure}[ht]
    % \vspace{-0.5cm}
    \includegraphics[width=\linewidth]{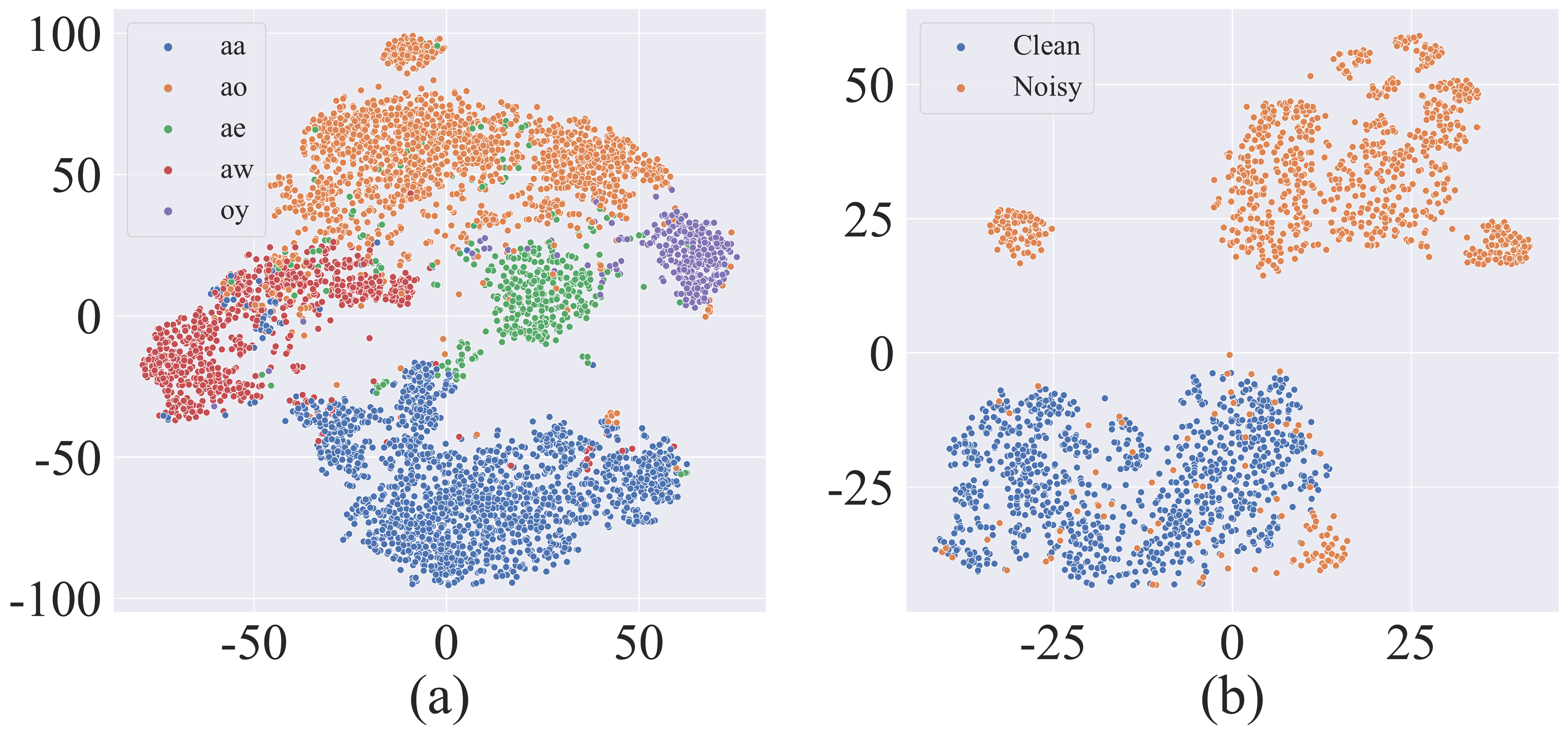}
    \centering 
    % \vspace{-0.2cm}
    \caption{t-SNE analysis on \textit{wav2vec} encoded feature map. (a) Feature map of five phone classes. (b) Feature map of clean and noisy utterances.}
    \label{fig:tsne}
    % \vspace{-0.4cm}
\end{figure}
\section{Experiments}
\label{sec:experiments}

In this section, we begin with the selected dataset and the evaluation metrics that were used as a standard benchmark. Next, we provide visualizations that demonstrate that the features generated by the PFPL are correlated with PESQ and STOI. Finally, it is shown that the proposed modification achieves competitive performance in terms of qualities and intelligibility.

% \begin{figure}[t]
%     % \vspace{-0.3cm}
%     \includegraphics[width=\linewidth]{figs/shift_sensitivity-2.pdf}
%     \vspace{-0.7cm}
%     \caption{Trending of different losses with its value reactions under different levels of time shift.}
%     \label{fig:shift}
%     \vspace{-0.4cm}
% \end{figure}

\subsection{Voice Bank--DEMAND Dataset}
\label{ssec:voice_bank_demand_dataset}
To compare our proposed SE system with other recent approaches, the Voice Bank--DEMAND dataset \cite{voicebank, demand} was used for evaluation. In this dataset, utterances recorded by 28 speakers out of a total 30 speakers were used for training, and the utterances from the remaining 2 speakers were used for testing. In the training set, noisy mixtures were synthesized using 10 types of noise at 4 different SNR levels, ranging from 0 dB to 15 dB, and 5 types of unseen noises, ranging from 2.5 dB to 17.5 dB were added to the testing set.

\begin{figure*}[t]
    \centering
    \includegraphics[width=0.87\textwidth]{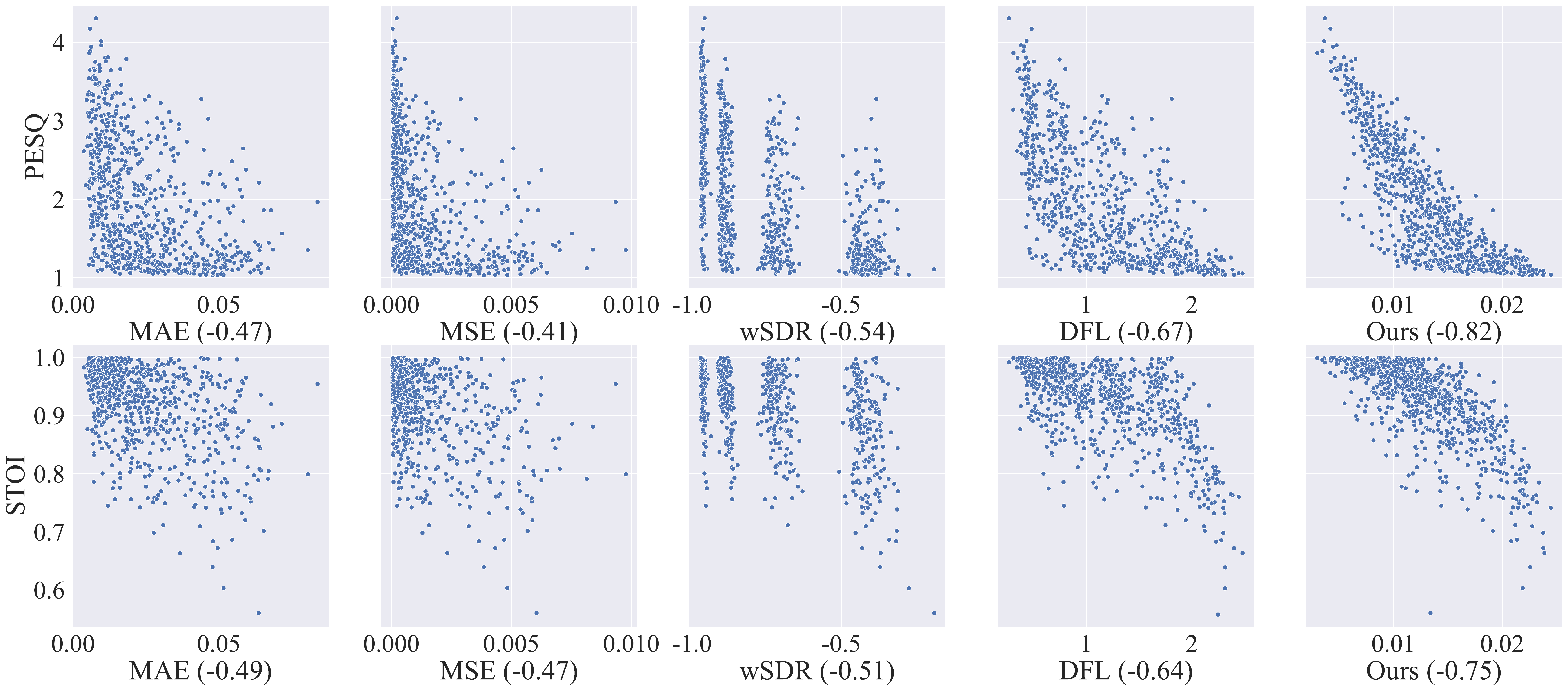}
    % \vspace{-0.2cm}
    \caption{Illustration the correlations of PESQ and STOI to different losses. To quantify how much a loss is correlated to a metric, we note the Pearson correlation coefficient in the parentheses. The higher absolute value of PCC indicates higher correlation.}
    % \vspace{-0.4cm}
    \label{fig:relation}
\end{figure*}

\subsection{Evaluation Metrics}
\label{ssec:evaluation_metrics}
Following prior works evaluated on the Voice Bank--DEMAND dataset, we used five metrics, which were CSIG, CBAK, COVL, introduced in \cite{metric}, PESQ, and STOI to measure the performance of the proposed method. CSIG, CBAK, and COVL demonstrate the signal distortion, background intrusiveness, and the overall quality with the same scale of mean opinion score, respectively. PESQ and STOI quantify the perceptual quality and the intelligibility of a speech signal. All of the above-mentioned metrics are better with higher scores.

% \subsection{Visualizing \textit{wav2vec} Features}
\subsection{Regarding SE as an Optimal Transport Problem}
\label{ssec:visulaization_on_wav2vec_features}
%As mentioned in Section \ref{ssec:problem_definition}, features extracted by \textit{wav2vec} are rich in phonetic information. In Fig. \ref{fig:tsne} (a), five types of phones that are properly separated, and therefore to some extent features generated by \textit{wav2vec} preserve the phonetic information. Fig. \ref{fig:tsne} (b) shows that most of the noisy and clean speech can be distinguished in the latent space. 
As mentioned in Section \ref{ssec:problem_definition}, latent representations of \textit{wav2vec} render rich phonetic information. Fig. \ref{fig:tsne}(a) demonstrates a t-SNE analysis of five phones, which are properly separated, confirming that the latent representations generated by \textit{wav2vec} carry rich phonetic information. Fig. \ref{fig:tsne}(b) shows that most of the noisy and clean speech are highly distinguishable in the latent space. Based on the observations from Fig. \ref{fig:tsne}, we can consider the training procedure of SE as an optimal transport problem that aims to search for a transformation mapping the distributions of noisy speech signals to that of the clean ones. Based on this concept, we decide to replace the $L^p$ distance and use the Wasserstein distance as the distance measure to compute the perceptual loss for the PFPL.
% meaning that the enhancement model is updated based on the phonetic distance between the estimation and its target.
% To verify that the generated features are competent at preserving phonetic information, we present an illustration that projects high-dimensional features into a two-dimensional space by t-SNE, which is a non-linear dimensionality reduction technique widely used for verifying the effectiveness of embeddings.

% \subsection{Shift Sensitivity}
% \label{ssec:shift_invariance}

% To evaluate the shift sensitivity of different losses, we illustrated the trending of losses over a time shift. Because losses have different scales, directly comparing loss change is impractical. Instead, we used the ratio of a loss to its second-smallest value, which is termed the discrepancy rate in Fig. \ref{fig:shift}. As shown in Fig. \ref{fig:shift}, PESQ is consistent at all levels of time shift. Comparing all the losses listed, DFL and our proposed PFPL are more insensitive than the other three signal-level losses, and more like auditory perception.

\begin{table}[t]
% \vspace{-0.3cm}
\begin{threeparttable}
\caption{Our proposed method versus some well performing methods with respect to different metrics. DFL\tnote{$\dag$} shows the results from the official source code and released parameters.}
% \cite{DFL:2019}
%\vspace{-0.3cm}
\centering
\small
\begin{tabular*}{\linewidth}{l|P{5.8mm} P{5.8mm} P{5.8mm} P{5.8mm} P{5.8mm}}
\hline
Model  & PESQ & CSIG & CBAK & COVL & STOI\\
\hline
\hline
\textbf{Noisy} & 1.97 & 3.35 & 2.44 & 2.63 & 0.92\\
\textbf{Wiener \cite{paliwal1987speech}} & 2.22 & 3.23 & 2.68 & 2.67 & --\\
\textbf{SEGAN \cite{pascual2017segan}} & 2.16 & 3.48 & 2.94 & 2.80 & 0.93\\
% \textbf{Wavenet} \cite{Rethage2017Wavenet} & - & 3.62 & 3.23 & 2.98 & -\\
\textbf{DFL \cite{germain2019speech}} & -- & 3.86 & 3.33 & 3.22 & --\\
\textbf{DFL}\tnote{$\dag$} & 2.58 & 3.80 & 2.72 & 3.19 & 0.93\\
\textbf{MetricGAN \cite{fu2019metricgan}} & 2.86 & 3.99 & 3.18 & 3.42 & 0.94\\
\textbf{HiFi-GAN \cite{su2020hifi}} & 2.94 & 4.07 & 3.07 & 3.49 & --\\
% \textbf{PHASEN} & 2.99 & 4.21 & 3.55 & 3.62 & 10.22\\
\textbf{SDR-PESQ \cite{kim2019end}} & 3.01 & 4.09 & 3.54 & 3.55 & --\\
\textbf{T-GSA \cite{kim2020t}} & 3.06 & {\bf 4.18} & 3.59 & 3.62 & --\\
\textbf{PERL-{\it wav2vec} \cite{kataria2020perceptual}} & 2.92 & 4.16 & 3.37 & 3.54 & 0.94\\
% \textbf{RDL--Net} & 3.02 & 4.38 & 3.43 & 3.72 & --\\
\hline
\hline
% \textbf{PFP}& 3.11 & 4.15 & 3.52 & 3.64 & 8.80\\
\textbf{PFP} & \textbf{3.15} & \textbf{4.18} & \textbf{3.60} & \textbf{3.67} & \textbf{0.95}\\
\hline
\end{tabular*}
\begin{tablenotes}
\scriptsize
\item[$\dag$] \href{https://github.com/francoisgermain/SpeechDenoisingWithDeepFeatureLosses.git}{https://github.com/francoisgermain/SpeechDenoisingWithDeepFeatureLosses.git}.
\end{tablenotes}
\label{tab:denoise}
% \vspace{-0.1cm}
\end{threeparttable}
\end{table}

\subsection{Correlation of Perceptual Metrics to Losses}
\label{ssec:relation_of_pesq_to_losses}
% A report of the full comparisons and analyses can be accessed on our GitHub page\footnote{\href{https://github.com/aleXiehta/PhoneFortifiedPerceptualLoss}{https://github.com/aleXiehta/PhoneFortifiedPerceptualLoss}}.
To analyze the relation between perceptual metrics and other losses, we compared several different losses to the corresponding metric scores on the testing set. Here, we illustrate the correlations of PESQ and STOI to five losses including, MAE, mean squared error (MSE), weighted source-to-distortion ratio (wSDR), DFL, and the proposed PFPL. Each point represents an utterance. From Fig. \ref{fig:relation}, we note that MAE and MSE have similar correlations to PESQ and STOI, and the four groups of points of wSDR represent the four SNR levels in the testing set. For the first three losses, there is no obvious correlation to the two metrics. However, the more obvious tendencies are that DFL and PFPL correlate to PESQ and STOI. Here, the Pearson correlation coefficient (PCC) is utilized to quantify the correlation between metrics and losses. The PCCs of losses are shown inside the parentheses in Fig. \ref{fig:relation}. The PCC of PFPL is much higher than all the other metrics' being compared. From Table \ref{tab:denoise} and Table \ref{tab:losses}, although DFL is more correlated with PESQ than the other three signal-level metrics, it has similar results to wSDR, MAE, and MSE. Because the PFPL measures how different the features are in terms of phonetic information salient to the human auditory system, it is reasonable that the PFPL is highly correlated with PESQ and STOI.
%\vspace{-0.2cm}
\subsection{Ablation Study on the PFPL}
\label{ssec:qualitative_analysis}
 In Table \ref{tab:denoise}, we compare prior approaches using GAN-based methods and specialized losses for auditory perception. Our approach achieved the highest PESQ score among all the compared methods. To understand PFPL, we compare several losses with the same model structure and conduct an ablation study on the PFPL. In Table \ref{tab:denoise}, {\bf PFPL}-$\mathcal{W}$ denotes {\bf PFPL} using the $L^p$ distance, and {\bf PFPL}-$\mathcal{W}$-MAE denotes {\bf PFPL}-$\mathcal{W}$ without using the MAE loss. From Table \ref{tab:losses}, point-wise losses (\textbf{wSDR}, \textbf{MSE}, and \textbf{MAE}) yield lower PESQ but higher CBAK comparing to the perceptual loss alone (i.e., {\bf PFPL}-$\mathcal{W}$-MAE). The low CBAK performance is caused by the point-wise difference ignored during training. This problem can be solved by adding MAE (i.e., the first term in Eq.~\eqref{eq:loss}) to the objective function, and accordingly {\bf PFPL}-$\mathcal{W}$ yields an improved CBAK score from 3.05 to 3.52. 
 Finally, by comparing {\bf PFPL} and {\bf PFPL}-$\mathcal{W}$,  the effect of the Wasserstein distance is confirmed, and our best results in terms of quality and intelligibility is attained by {\bf PFPL}.

\begin{table}[t]
\caption{Comparison of the ablations of PFPL and the point-wise losses with respect to evaluation metrics.}
% \vspace{0.3cm}
\centering
\small
\begin{tabular*}{\linewidth}{l|P{7.2mm} P{7.2mm} P{7.2mm} P{7.2mm} P{7.2mm}}
\hline
Loss & PESQ & CSIG & CBAK & COVL & STOI\\
\hline
\hline
\textbf{wSDR \cite{choi2018phase}} & 2.58 & 3.00 & 3.18 & 2.76 & 0.93\\
\textbf{MSE} & 2.60 & 3.31 & 3.19 & 2.94 & 0.93\\
\textbf{MAE} & 2.62 & 3.47 & 3.20 & 3.02 & 0.93\\
\hline
\hline
\textbf{PFPL}-$\mathcal{W}$-MAE & 3.09 & \textbf{4.22} & 3.05 & \textbf{3.67} & 0.94\\
\textbf{PFPL}-$\mathcal{W}$ & 3.11 & 4.15 & 3.52 & 3.64 & \textbf{0.95}\\
\textbf{PFPL}& \textbf{3.15} & 4.18 & \textbf{3.60} & \textbf{3.67} & \textbf{0.95}\\
\hline
\end{tabular*}
% \vspace{-0.1cm}
\label{tab:losses}
\end{table}

\section{Conclusion}
\label{sec:conclusion}
%In this paper, based on prior works on perceptual loss functions, we proposed an improved perceptual loss that uses Wasserstein distance to replace the $L^p$ distance in the original perceptual loss.
In this paper, we have proposed a novel PFPL loss for training SE models. The PFPL is derived based on the latent representations of the {\it wav2vec} model, which carry rich phonetic information. Meanwhile, the PFPL uses the Wasserstein distance as the distance measure. Accordingly, the SE training can be seen as an optimal transport problem that aims to move the latent representation distributions of noisy speech to that of clean speech. The experimental results first revealed that the PFPL has very high correlations with perceptual metrics as  compared to other related loss functions. Moreover, the SE model trained with the PFPL outperforms several well-known and related works in terms of standardized evaluation metrics. 
% Additionally, to tackle the shortage of perceptual loss, we regulated perceptual by MAE, so the enhancement model not only learned to distinguish latent representations, but also learned in time domain, and the disruption by background noise was notably eliminated with the modification. 
\vfill\pagebreak

\bibliographystyle{IEEEtran}

\bibliography{mybib}

% \begin{thebibliography}{9}
% \bibitem[1]{Davis80-COP}
%   S.\ B.\ Davis and P.\ Mermelstein,
%   ``Comparison of parametric representation for monosyllabic word recognition in continuously spoken sentences,''
%   \textit{IEEE Transactions on Acoustics, Speech and Signal Processing}, vol.~28, no.~4, pp.~357--366, 1980.
% \bibitem[2]{Rabiner89-ATO}
%   L.\ R.\ Rabiner,
%   ``A tutorial on hidden Markov models and selected applications in speech recognition,''
%   \textit{Proceedings of the IEEE}, vol.~77, no.~2, pp.~257-286, 1989.
% \bibitem[3]{Hastie09-TEO}
%   T.\ Hastie, R.\ Tibshirani, and J.\ Friedman,
%   \textit{The Elements of Statistical Learning -- Data Mining, Inference, and Prediction}.
%   New York: Springer, 2009.
% \bibitem[4]{YourName17-XXX}
%   F.\ Lastname1, F.\ Lastname2, and F.\ Lastname3,
%   ``Title of your INTERSPEECH 2021 publication,''
%   in \textit{Interspeech 2021 -- 20\textsuperscript{th} Annual Conference of the International Speech Communication Association, September 15-19, Graz, Austria, Proceedings, Proceedings}, 2020, pp.~100--104.
% \end{thebibliography}

\end{document}